\documentstyle{article}

\def\mod{~{\rm mod}~}

\title{\begin{flushright}
\small June 1998 \hfill SINP/TNP/98-14\\
{\tt hep-ph/9806274} 
\end{flushright}
{\bf A reincarnation of $R$-parity}}

\author{
\bf Palash B. Pal%
\thanks{E-mail address: pbpal@tnp.saha.ernet.in} \\ 
{\em Saha Institute of Nuclear Physics, 1/AF Bidhan Nagar, 
Calcutta 700064, India}
}

\date{}

\begin{document}

\maketitle

\begin{abstract}

{\small In supersymmetric theories, $R$-parity is defined in a way
such that it does not commute with the space-time symmetries. We show
that, in general sypersymmetric models, 
one can define a discrete symmetry which commutes with all
space-time and gauge symmetries, and whose phenomenological
implications are equivalent to those of $R$-parity.}

\end{abstract}

In supersymmetric field theories, $R$ symmetry \cite{U1R} is a general
class of symmetries under which the fermionic co-ordinate of the
superspace transforms non-trivially. Among these, the discrete
$R$-parity \cite{Rpar} has proved to be an important tool in the
analysis of supersymmetric gauge theories, in particular the minimally
supersymmetric standard model (MSSM). It is defined to be a discrete
$Z_2$ symmetry under which any particle has the quantum number
	\begin{eqnarray}
R = (-1)^{3(B-L)+2S} \,,
	\end{eqnarray}
where $B$ and $L$ are the baryon and lepton numbers, and $S$ is the
spin of the particle. Thus, under this, all ordinary particles in the
standard model are even, whereas all of their superpartners are
odd. In a theory with $R$ invariance, then, superpartners can be
produced or annihilated only in pairs.

The $R$-parity does more than that. The most general superpotential of
the MSSM which is consistent with gauge symmetry and supersymmetry can
be written as
	\begin{equation} 
W = W_0 + W',
	\end{equation} 
where, using the usual notation for the superfields, $W_0$ and $W'$
are given by~\cite{WforMSSM}
	\begin{eqnarray}
\label{W0}
W_0 & = & f_e^{ij} L_i H_d E_j^c + f_d^{ij} Q_i H_d D_j^c
+ f_u^{ij} Q_i H_u U_j^c + \mu H_d H_u \,, \\
\label{W'}		
W' & = & {1\over{2}}\lambda_{ijk} L_i L_j E^c_k + 
\lambda'_{ijk} L_i Q_j D^c_k + 
{1\over{2}}\lambda''_{ijk} U^c_i D^c_j D^c_k + 
\mu_i  L_i H_u \,.
	\end{eqnarray}
The terms in $W'$ violated either baryon number or lepton number and
can mediate $B$ and $L$ violating processes at a huge rate unless the
coupling constants are very small \cite{review}. However, when $R$
parity is imposed 
on the theory, these terms cannot appear in the superpotential, which
is an elegant way of making the theory phenomenologically acceptable.

The awkwardness with the $R$-parity is that its definition includes
the spin of the particle, so that the symmetry does not commute with
space-time supersymmetry. This is manifest by the fact that an
ordinary particle and its superpartner, which belong to the same
supermultiplet, have opposite $R$-parity assignments.

It is of course possible to rule out the terms in $W'$ by imposing a
different discrete symmetry. For example, consider the $Z_2$ symmetry
under which all the superfields containing quarks and leptons change
sign, whereas those containing the Higgs and the gauge bosons do not
\cite{DG}. This certainly prohibits all the $B$ and $L$ violating
terms present in $W'$, although it is not obvious whether it implies
that the superpartners are always produced in pairs. On the other
hand, it is obvious that such a symmetry commutes with all space-time
and gauge symmetries, because the transformation is the same on all
component fields in any supermultiplet.

What we want to show in this note is the equivalence of these two
types of symmetries in a class of $N=1$ supersymmetric models
containing the MSSM and almost all its extensions. To define this
class as well as to set up the notation, let us divide all superfields
in a model into two sets which we will denote by $F$ and $B$. The
names of these two sets are suggestive of the fact that we make a
classification of the component fields in which the fermionic
components of the superfields in $F$ as well as the bosonic components
of the members of $B$ to be ``ordinary fields'', whereas the
complementary fields are called ``superpartners''. If we denote
fermionic components of superfields by the corresponding lower case
letters and the bosonic components by script capitals, the component
fields $f$ and $\cal B$ are ordinary, whereas the components $\cal F$
and $b$ are superpartners. We summarize the notation in a tabular
form for future reference:
	\begin{eqnarray}
\begin{tabular}{l|cc}
& Fermion & Boson \\ \hline
Ordinary particle & $f$ & $\cal B$ \\ 
Superpartner & $b$ & $\cal F$ \\ 
\end{tabular}
	\end{eqnarray}
The assumption of the model is that the sets
$F$ and $B$ are disjoint, i.e., no ordinary particle is the
superpartner of another ordinary particle. This assumption certainly
holds for the conventional classification of the fields in MSSM, in
which the quarks and leptons, the Higgs and the gauge bosons are
called ordinary. But we emphasize that the result that we are going to
prove is true for any assignment of the component fields into ordinary
fields and superpartners as long as the disjointness criterion is
satisfied.

We now define a generalized $R$-parity as a $Z_2$ symmetry which
guarantees that the superpartners are produced or annihilated in
pairs. This is guaranteed by a symmetry under which all superpartners
change sign, whereas the ordinary fields do not. In other words, the
component fields transform as follows: 
	\begin{eqnarray}
\begin{tabular}{c|cccc}
Component field & $f$ & ${\cal F}$ & ${\cal B}$ & $b$ \\ \hline 
$R$ eigenvalue & $+$ & $-$ & $+$ & $-$ \\
\end{tabular}
\label{R}
	\end{eqnarray}
As commented earlier, this symmetry does not commute with the
space-time symmetries. Let us now consider another symmetry of the
type mentioned above for the MSSM. We call it the $A$-parity. Under
this, the eigenvalues of different fields are given below:
	\begin{eqnarray}
\begin{tabular}{c|cccc}
Component field & $f$ & ${\cal F}$ & ${\cal B}$ & $b$ \\ \hline 
$A$ eigenvalue & $-$ & $-$ & $+$ & $+$ \\
\end{tabular}
\label{S}
	\end{eqnarray}
Alternatively, we can say that under this symmetry operation, the
superfields $F$ change sign, whereas the superfields $B$ do not. Thus,
this symmetry commutes with space-time symmetries.

We now show that all phenomenological consequences of these two
symmetries are equivalent. For this, consider a generalized operator
in the component field notation, which we write as
	\begin{eqnarray}
f^{n_f}  {\cal F}^{n_{\cal F}}  {\cal B}^{n_{\cal B}}  b^{n_b} \,.
\label{genop}
	\end{eqnarray}
Let us assume that our theory is $R$ invariant. Then, using Eq.\
(\ref{R}), we find that the powers of different fields must satisfy
the relation
	\begin{eqnarray}
n_{\cal F} + n_b = 0 \mod 2 \,.
	\end{eqnarray}
In addition, the operator must be Lorentz scalar. This requires that
there is an even number of fermionic fields in the generalized
operator of Eq.\ (\ref{genop}). Since the component
fields $f$ and $b$ are fermionic, this implies
	\begin{eqnarray}
n_f + n_b = 0 \mod 2 \,.
	\end{eqnarray}
Adding these two conditions, we obtain
	\begin{eqnarray}
n_f + n_{\cal F} = 0 \mod 2 \,,
	\end{eqnarray}
which is the condition imposed on the operator in Eq.\ (\ref{genop})
from $A$-parity. Thus, $R$-parity implies $A$-parity. Exactly
similarly, we can show that $A$-parity also implies $R$-parity.
Thus we have shown that any operator which is not allowed by
$R$-parity is also not allowed by $A$-parity, and vice versa.
This is the general result.

A specific example might help understand the equivalence further. We
know that an $R$ invariant theory implies that the lightest
superpartner (LSP) will be stable. This is obvious from the
$R$-assignments in which all superpartners are negative. Thus, one
superpartner cannot decay into ordinary particles which are all
positive under $R$-parity. From $A$-parity assignments, this is not as
obvious to see. But it is nevertheless true. To see this, let us first
deal with the possibility of two-body decay modes. Suppose the LSP is
a fermion, i.e., belongs to the class $b$ in our notation. It will
then have to decay into an ordinary fermion and an ordinary boson,
i.e., to a combination $f{\cal B}$. But $b$ and $\cal B$ are even
under $A$, whereas $f$ is odd. So this is not possible. Similarly, if
the LSP is a boson, it will have to decay either to a combination $ff$
or to ${\cal BB}$. Both are impossible since the bosonic
superpartners, which we called $\cal F$, are odd under $A$. The
arguments can be easily extended to consider more than two particles
in the final state.

Similarly, we can show that the imposition of $A$-parity prohibits
all baryon and lepton number violating renormalizable terms in the
Lagrangian of MSSM. However, in the MSSM, the $A$-parity can be
identified as 
	\begin{eqnarray}
A = (-1)^{3(B-L)} \,.
\label{Adef}
	\end{eqnarray}
From this, it is tempting to conclude that imposing $A$-parity is
equivalent to imposing $B-L$ as a global symmetry. But this would be
an unfair conclusion for many reasons. First, $B-L$ is a continuous
U(1) symmetry, whereas $A$ is a discrete symmetry. We are employing a
smaller symmetry to obtain a larger symmetry on the renormalizable
interactions. Second, we have proved our result in a more generalized
context, where we need not follow the standard classification of the
particles into ``ordinary'' and ``superpartners'', and $A$ need not be
defined as in Eq.\ (\ref{Adef}).

$R$-parity is a symmetry which does not commute with space-time
symmetries. We have defined a generalized $A$-parity which does. And
we have also shown that the consequences of these two symmetries are
identical. We feel that in this case, it is more convenient to talk
about the $A$-parity rather than the $R$-parity. For example, the
$R$-invariant MSSM can be called a supersymmetric model based on the
symmetry ${\rm SU(3)}_c \times {\rm SU(2)}_L \times {\rm U(1)}_Y
\times A$, under which, for example, the leptonic doublet $L_i$
transforms as $(1,2,-{1\over2})_-$ and the gluon superfield 
as $(8,1,0)_+$, where the subscripted signs denote the $A$
eigenvalue. Besides, in theories like grand unified theories, it is
much easier to define $A$-parity than $R$-parity since the latter is
defined through $B$ and $L$ quantum numbers which are not defined in
the gauge interactions of most grand unified models.

I thank Gautam Bhattacharyya for enlightening and stimulating
discussions.


\begin{thebibliography}{[W]}

\bibitem{U1R} A. Salam, J. Strathdee: Nucl. Phys. B87 (1975) 85;\\
P. Fayet: Nucl. Phys. B90 (1975) 104.

\bibitem{Rpar} G. Farrar and P. Fayet, {Phys. Lett.} {B76}
(1978) 575.

\bibitem{WforMSSM} S. Weinberg, {Phys. Rev.} {D26} (1982)
287;\\ 
N. Sakai and T. Yanagida, {Nucl. Phys.} {B197} (1982) 533.


\bibitem{review} For recent reviews and references, see, e.g.,\\ 
G. Bhattacharyya, {\tt hep-ph/9709395} and 
Nucl. Phys. Proc. Suppl. {52A} (1997) 83;\\ H. Dreiner: {\tt
hep-ph/9707435}.

\bibitem{DG} S. Dimopoulos, H. Georgi: Nucl. Phys. B193 (1981) 150;\\ 
G. Costa, F. Feruglio, F. Zwirner: Nuovo Cimento A70 (1982) 201.

\end{thebibliography}
\end{document}